\documentclass[final,3p]{elsarticle}

\usepackage[colorlinks=true,breaklinks=true,pdftex]{hyperref}
\usepackage{times}
\usepackage{epsfig}
\usepackage{graphicx}
\usepackage{amsmath}
\usepackage{amssymb}
\usepackage{subfigure}%
\usepackage{overpic}
\usepackage{graphicx}
\usepackage{gensymb} 
\usepackage{amsmath,amsfonts,amssymb}
\usepackage{overpic}
\usepackage{contour}\contourlength{0.7pt}
\usepackage{algorithm}
\usepackage{algpseudocode}

\usepackage[colorlinks=true,breaklinks=true,pdftex]{hyperref}

\biboptions{authoryear}
 
\begin{document}
 
\bibliographystyle{plain} 

\begin{frontmatter}
 
\title{Developable B-spline surface generation from control rulings}
 
 \author[first]{Zixuan Hu } 
 \author[first]{Pengbo Bo \corref{cor}} \ead{pbbo@hit.edu.cn}
 \cortext[cor]{Corresponding author} 
\address[first]{School of Computer Science and Technology, Harbin Institute of Technology, Weihai 264209, China}
 
\begin{abstract}
An intuitive design method is proposed for generating developable ruled B-spline surfaces from a sequence of straight line segments indicating the surface shape. The first and last line segments are enforced to be the head and tail ruling lines of the resulting surface while the interior lines are required to approximate rulings on the resulting surface as much as possible.  This  manner of developable surface design is conceptually similar to the popular way of the freeform curve and surface design in the CAD community, observing that a developable ruled surface is a single parameter family of straight lines. This new design mode of the developable surface also provides more flexibility than the  widely employed way of developable surface design from two boundary curves of the surface.  The problem is treated by numerical optimization methods with which a particular level of distance error is allowed. We thus provide  an  effective tool for creating surfaces with a high degree of developability  when the input control rulings do not lie in exact developable surfaces.  We consider this ability as the superiority over analytical methods in that it can deal with arbitrary design inputs and  find  practically useful results. 
\end{abstract}
 

\end{frontmatter}

\section{Introduction and motivation}
In industrial manufacturing including the ship-hull building and free-form architecture building, the employed materials,  such as sheet metal and wooden panels, often possess  the physical property of high bendability and low stretchability. The developable surface is a suitable mathematical model for these materials and thus has important industrial applications~\cite{Baldassini2008Freeform}~\cite{Julie1998Design}~\cite{PEREZ2007853}. Developable surfaces have also been investigated in CNC flank milling~\cite{Chu2005ToolPP}~\cite{Calleja2018HighlyA5}. However, the current commercial CAD software do not have flexible and effective capabilities for modeling developable surfaces.  The study of design methods for developable surfaces has been an active research topic in recent years and a variety of approaches have been proposed~\cite{Tang2016Interactive}~\cite{Ming2011Design}~\cite{Caiyun2020Designing}~\cite{Alicia2015Interpolation}. 

\begin{figure}[htb]
\centerline{
\scriptsize
\hfill
\begin{minipage}[b]{0.35\linewidth}
\centering
\begin{overpic}[width=0.8\textwidth]{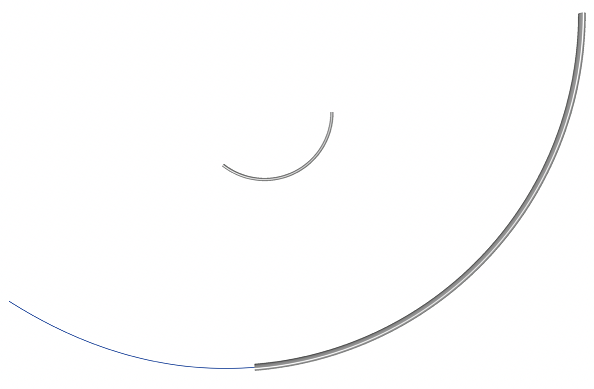}
\end{overpic}\\
\centering (a)
\end{minipage}
\hfill
\begin{minipage}[b]{0.3\linewidth}
\includegraphics[width=0.8\textwidth]{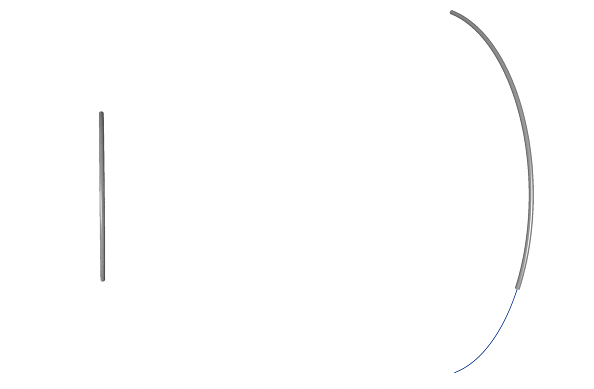}\\
\centering  (b)
\end{minipage}
\hfill
\begin{minipage}[b]{0.3\linewidth}
\includegraphics[width=0.8\textwidth]{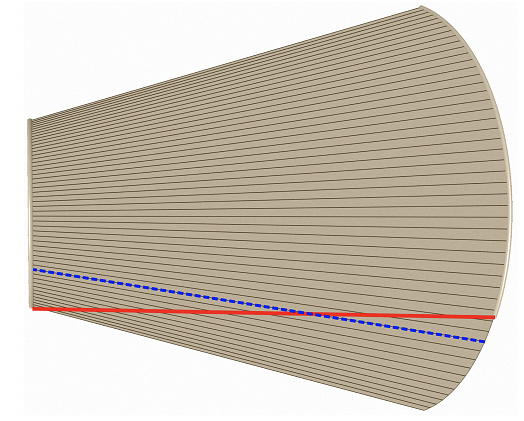}\\
\centering  (c)
\end{minipage}
\hfill
}
\caption{Developable surface modeling based on given design curves. (a)(b) Two different perspectives of two initial design curves (thick curves) and the extension line of one design curve (thin curve). (c) Developable surface bounded by two given curves. The red line is the trimming line, and the blue dotted line is one of the rulings to be trimmed.}
\label{fig:fig1}
\end{figure}

To be compatible with the surface representation in commercial CAD software, the designed  surface is required to be represented by a B-spline surface. With the common design method of developable B-spline surfaces, the surface shape is expressed by two curves, denoted by $C_0(t)$ and $C_1(T)$, $t$, $T\in[0,1]$, serving as surface boundaries and  a  (quasi-) developable surface bounded by $C_0(t)$ and $C_1(T)$ (or their perturbation curves) is constructed. The resulting surface generally does not take the connection lines of the endpoints of the input curves (i.e. $\overline{C_0(0)C_1(0)}$ and $\overline{C_0(1)C_1(1)}$) as its ruling lines. Therefore, the  input curves need to be adjusted through curve extension, and  the generated surface needs to be trimmed against the terminal connection lines~\cite{bo2019multi}. Fig.~\ref{fig:fig1} shows a simple example of developable surface design based on boundary curves, which illustrates the necessity of curve extension and surface trimming. 

\begin{figure}[t]
\centerline{
\scriptsize
\hfill
\begin{minipage}[b]{0.27\linewidth}
\centering
\begin{overpic}[width=0.9\textwidth]{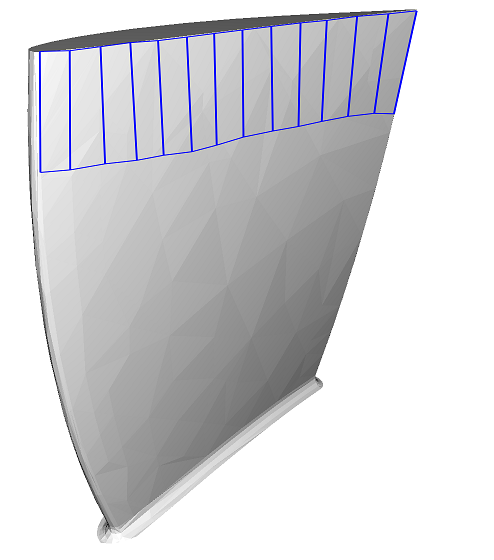}
\end{overpic}\\
\centering (a)
\end{minipage}
\hfill
\begin{minipage}[b]{0.25\linewidth}
\centering
\begin{overpic}[width=0.9\textwidth]{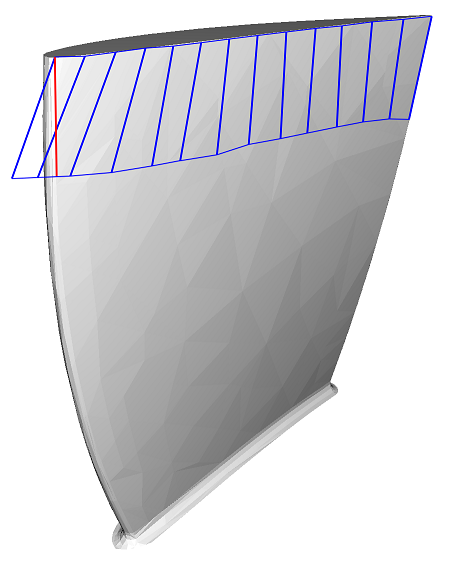}
\end{overpic}\\
\centering (b)
\end{minipage}
\hfill
}
\caption{The ruling lines on a blade surface.  (a) The edge of the blade aligns with the  ruling lines of the ruled surface. (b) The edge of the blade becomes the trimming line (the red line) of the ruled surface.}
\label{fig:BladeRuling}
\end{figure}

This extension-trimming method has some drawbacks: 1) The developability of the resulting surface largely depends on the shape of the extension curves and it is not clear how to extend the original curves in favour of surface developability. Therefore,  the shape of extended curves cannot be intuitively estimated, which may draw a big difference between the result and the original design intention. 2) In some applications, it is desired that the  developable surface is bounded by two ruling lines instead of trimming lines. For example, a blisk blade surface is often represented by a ruled surface and it will facilitate the process of CNC flank milling with conical or cylindrical cutters if the blade surface is bounded by ruling lines without trimming. In Fig.~\ref{fig:BladeRuling}, the design of the ruled surface with and without edge trimming of a blade model are shown.  In the case of being bounded by ruling lines (Fig.~\ref{fig:BladeRuling}(a)), the two side surfaces of the blade can be connected by a transition rounded surface which is also ruled, and consequentially can  be flank-milled in one pass. However, for the design in Fig.~\ref{fig:BladeRuling}(b), the transition part has to be point-milled in a separate operation, adding production time.  Therefore, the tool for designing a ruled surface with specified terminal rulings are highly demanded in applications.

To provide an intuitive design method, we treat developable surface modeling following the traditional design process of  curves and surfaces. In terms of curve design, the designer first describes the shape of the curve by defining some shape-control points and then constructs a curve fitting the shape-control points. In many cases, the resulting curve is required to take the first and last shape-control points as its endpoints, namely, the resulting curve needs to strictly interpolate the first and last shape-control points, while trying to fit the interior shape-control points. Currently, this intuitive and friendly design method is the main design method of free-form curves and surfaces, but similar techniques are not available in developable surface design.

The developable surface is the envelope  of a single-parameter family of planes and   designing developable surfaces through a set of planes has been considered in some works. These methods working with the dual representation of a developable surface and are lack of design intuition and stability~\cite{bodduluri1993design}. Instead, we work directly in the design space and treat  a  developable ruled surface as a single parameter family of lines. We propose to describe the shape of a developable surface by  a sequence of straight line segments and aim to generate a developable surface taking the input lines as its ruling lines. More precisely,  we  construct a developable B-spline surface which  takes  the first and last input lines as the head and tail rulings, respectively, and exhibits  the shape suggested by  the interior lines as much as possible. The main contributions of our work are summarized as follows.

\begin{itemize}
\item Our method works directly in the design space by specifying a sequence of control rulings and is thus geometrically intuitive for developable surface design.
\item Different to existing methods which model a developable surface from its boundary curves, our method allows the user to specify the first and the last rulings explicitly.
\item By treating the terminal control rulings as hard constraints and interior control rulings as soft constraints, our method is capable of computing a B-spline surface with a high degree of developability.
\end{itemize}

\section{Related works}
Developable surface modeling has been extensively studied and  various representations of the developable surface have been considered. In computer animation and  simulation, the triangular and quadrilateral meshes are most frequently used. For the works on developable surface modleing based on quadrangular and triangular meshes, refer to ~\cite{English2008Animating}~\cite{Solomon2012Flexible}~\cite{Liu2006Geometric}~\cite{LIU20111089}. In CAD community, B-spline/NURBS surfaces are the most frequently used representations~\cite{PEREZ2007853}~\cite{Oetter2002Block}. Subag and Elber approximate a general NURBS surface using piecewise developable surfaces with a global error bound ~\cite{Subag2006Piecewise}. Pottmann et al. use the composite developable strip model to obtain free-form surfaces in architecture and manufacturing~\cite{Baldassini2008Freeform}.

A widely studied problem in developable surface modeling is to construct a developable surface bounded by two given design curves. In ship-hull design, the hull shape is described by some feature curves, and then the developable surface is constructed between two adjacent feature curves~\cite{PEREZ2007853}. Tang and Wang simulate the folding process of elastic sheets to approximate a developable surface  via the operation of boundary triangulation~\cite{Kai2006Modeling}. Wang et al. formulate the developable triangulation problem as a graph problem and use the Dijkstra algorithm to solve it ~\cite{Wang2005Optimal}~\cite{Tang2009Quasi}. A local-global method is proposed to improve surface developability by optimizing mesh vertices~\cite{Ming2011Design}. The surfaces obtained by these methods are discrete mesh surfaces.

The smooth developable ruled surfaces are often represented as the B\'{e}zier or  the B-spline surfaces. It is required that the surface satisfies the nonlinear developability constraint  and particular boundary conditions (such as fixing one  boundary curve). Some works study the analytical computation of developable surfaces from the nonlinear developability equations~\cite{PottmannDevelopable}~\cite{CHEN1999110}~\cite{AUMANN2003601}~\cite{Aumann:2004:0167-8396:661}~\cite{Caiyun2020Designing}~\cite{hu2019Generalized}. Chu study the degrees of freedom in a developable B-spline surface bounded by two curves~\cite{Chu2013Geometric}. The main issue with analytical methods is that  the given inputs are required to define precise developable surfaces and this cannot be guaranteed in real applications such as ship-hull design and free-form architectural design. Chen and Tang generate a developable surface by assembling a set of smooth surface patches. Triangular B\'{e}zier patches are used for creating smooth surfaces with $G^1$ continuity~\cite{ChenQuasi} while quadrilateral B\'{e}zier patches are used for constructing smooth surfaces with $G^2$ continuity~\cite{ChenG2}. However, these methods are not capable of constructing a developable surface strictly bounded by given curves. Moreover, the resulting surfaces  do not possess explicit ruling lines  which provide important guiding information in the real manufacturing process. Bodduluri et al. make use of the  duality between plane and point geometries for developable surface modeling.  However, this method is weak in terms of geometric intuition ~\cite{bodduluri1993design}. Although analytical derivation of precise developable surfaces is mathematically interesting, they can not handle arbitrary inputs with which only approximate results can be obtained. 

Numerical  methods for developable surface design have been widely studied in recent years. The goal is to construct a quasi-developable surface that meets application error tolerance. Tang et al. propose an interactive design method and decrease the degree of the constraint equations by introducing auxiliary variables~\cite{Tang2016Interactive}. P\'{e}rez et al. study the application of quasi-developable B-spline surfaces in ship-hull design and use the multi-conic method to modify the given curves to improve surface developability~\cite{PEREZ2007853}. A method has been proposed for computing a developable surface bounded by  curves perturbed from original design curves, but the resulting surface is not a B-spline surface~\cite{bo2019multi}.  Bo et al. propose a method for constructing a quasi-developable B-spline surface between given B-spline curves, and the resulting surface is strictly interpolated to given boundary curves~\cite{bo2020asdevelopable}. As we have explained, the  design method based on  boundary curves is not a reasonable way to control the terminal rulings of the surface.   

Ruling lines are considered as design guidance in some works.  Chalfant et al. propose a quasi-developable design method based on given boundary curves and boundary rulings, and discuss its applications in  ship-hull surface modeling~\cite{Julie1998Design}. Park at el. give the direction of a set of ruling lines and two endpoints of the boundary curve, and obtain developable surfaces by optimal control~\cite{Junghyun2002Design}. Fernandez-Jambrina designs a developable surface by giving one boundary curve and two boundary rulings, but the endpoints of the boundary rulings cannot be specified by the designers~\cite{FernB}. Caton et al. improve the method in ~\cite{FernB} by  the degree elevation operation and provide the ability of choosing both endpoints of the rulings~\cite{Alicia2015Interpolation}. These methods generate accurate developable surface  and   have strict requirements on boundary conditions. Therefore, these methods are not suitable for solving practical problems where the input curves or ruling lines may not contribute to a precise developable surface.  In this paper, we propose an intuitive design method based on numerical computation for constructing a quasi-developable surface by  specifying some ruling lines  freely in space as control tools.

\section{Developable surface design through control rulings}

According to the above discussions, the problem we study in this paper is defined as follows. Given  in space  a set of ordered line segments $L_i=(Q_i,P_i),i=0,...,K$, where $Q_i,P_i$ are the two endpoints of $L_i$, the objective is to construct a developable B-spline surface $S$, defined by  

\begin{equation}
S(t,s)=C_0(t)(1-s)+C_1(t)s, \quad t,s\in[0,1]
\label{eq:eq1}
\end{equation}

\noindent where $C_0(t)$ and $C_1(t)$ are B-spline curves with the clamped knot vectors.  $S$ is required to  take the line segments $L_i$, $i=0,...,K$ as its ruling lines as much as possible with the first and the last line segments being its terminal rulings, i.e.,  to satisfy $S(0,s)=L_0$ and $S(1,s)=L_K$. In this way, the input line segments  serve as tools for controlling the shape of the developable surface and are thus called the \emph{control rulings}.

Fitting  $S$  to the rulings $L_i,i=0,...,K$ can be expressed by  fitting the  curves $C_0(t)$ and $C_1(t)$ to the corresponding endpoints of the rulings, i.e.,  to satisfy  $C_0(t_i)=Q_i$ and $C_1(t_i)=P_i,i=0,...,K$, where $t_i$ are the parametrization of the input ruling lines. It is clear that if we have  $C_0(t_i)=Q_i$ and $C_1(t_i)=P_i$, the ruling line $S(t_i,s)$ on the surface is identical to the  input control ruling connecting $Q_i$ and $P_i$, denoted by $\overline{Q_iP_i}$. Therefore, the problem of developable surface generation is transformed into the problem of curve fitting, namely, finding two B-spline curves $C_0(t)$, $C_1(t)$ fitting the data points $Q_i$ and $P_i,i=0,...,K$, respectively. Different to  a general curve fitting problem,  the curves $C_0(t)$ and $C_1(t)$ are interrelated in  developable surface modeling by the following constraints.

\begin{itemize}
\item  Constraint 1: The data points $Q_i$ and $P_i$ correspond to the same parameter value $t_i$.
\item  Constraint 2: The surface $S$ bounded by $C_0(t)$ and $C_1(t)$, defined by Eq.(\ref{eq:eq1}), achieves a high degree of developability.
\end{itemize}

Because there are generally no accurate developable surfaces interpolating arbitrary line sequences, we have to make some relaxations and compute developable surfaces numerically. Moreover, the input ruling lines do not have equal significance in design. In our method, we treat the first and the last control rulings as strict interpolation constraints and the interior control rulings as soft constraints. In the following, we propose some specific design manners using the control rulings and discuss the capabilities of various design ways.

\subsection{\textbf{Developable surface from control rulings and one fixed boundary curve}}
\label{sec:onefixboundary}
In some applications, one of the boundary curves of the developable surface is given and fixed. Given  additionally a set of straight line segments serving as ruling lines, we need to construct the other boundary curve to obtain the developable surface $S$. This problem is specifically defined as follows. Suppose we are  given a  boundary curve, denoted by $C_0(t), t\in[0,1]$, and a sequence of line segments $L_i(i=0,...,K)$ emanating from  points on $C_0(t)$.  The problem of developable surface computation from the control rulings can be described by the constraints:  the line segments $L_i$ are identical to $\overline{C_0(t_i)P_i}$, $i=0,...,K$, with $t_0=0,t_K=1$ indicating the terminal interpolation constraint and $P_i$ being the  data points in space.

As we have explained, this ruling fitting problem can be solved by curve fitting with particular constraints. The unknown variables  are the control parameters of   $C_1(t)$ whose endpoints are  fixed to be  $P_0$ and $P_K$. The objective is to make   $C_1(t)$   approximate $P_i, i = 1, ..., K-1$ with which the surface $S$ defined in Eq.(\ref{eq:eq1}) achieves a high degree of developability.

For arbitrary input rulings, there is generally no developable surface exactly interpolating the  rulings. However, a B-spline surface with a large  number of control points can provide enough degrees of freedom to construct a quasi-developable surface. We formulate this problem as a numerical optimization problem and  obtain the resulting curve $C_1(t)$ and the developable surface $S$ by minimizing an objective function evaluating surface developability.

\emph{\textbf{A prime-dual formulation of developability}}. The surface $S$ (Eq.(\ref{eq:eq1})) bounded by $C_0(t)$ and $C_1(t)$ being a developable surface requires $S$ to meet the following constraints~\cite{Tang2016Interactive}.

\begin{equation}
\left\{
             \begin{array}{lr}
             C_0^{'}(t) \cdot N(t)=0 &  \\
             C_1^{'}(t) \cdot N(t)=0 & \\
             (C_0(t)-C_1(t)) \cdot N(t)=0 &  
             \end{array}
\right.
\nonumber
\end{equation}

\noindent where $N(t)$ is a B-spline function representing the normal vector field of the surface. Using the normal vector function comprehensively combines the prime and dual form of the developable surface, which leads to a  stable convergence with our optimization. Moreover, the normal function $N(t)$ helps to regularize the surface owing to the inherent smoothness of a B-spline curve. However, using a B-spline function for the normal vector field also means that all normal vectors are linear combinations of the same set of control points and this reduces the degrees of freedom  for improving surface developability. Therefore, we adopt independent normal vectors $N_i$ at a set of samplings instead of a smooth B-spline function for the normal vector field.  We thus introduce the following term to evaluate surface developability.

\begin{align}
F_{Dev} \equiv \sum_{k=0...M}(C_0^{'}(t_k) \cdot N_k)^2+(C_1^{'}(t_k) \cdot N_k)^2+((C_0(t_k)-C_1(t_k)) \cdot N_k)^2=0
\label{eq:eq2}
\end{align}

\noindent where $t_k$ are sampling parameters. $N_k$ are the normal vectors corresponding to  $t_k$ which are also the variables in optimization, in addition to the interior control points of $C_1(t)$.

\begin{figure}[t]
\centerline{
\scriptsize
\hfill
\begin{minipage}[b]{0.5\linewidth}
\centering
\begin{overpic}[width=0.7\textwidth]{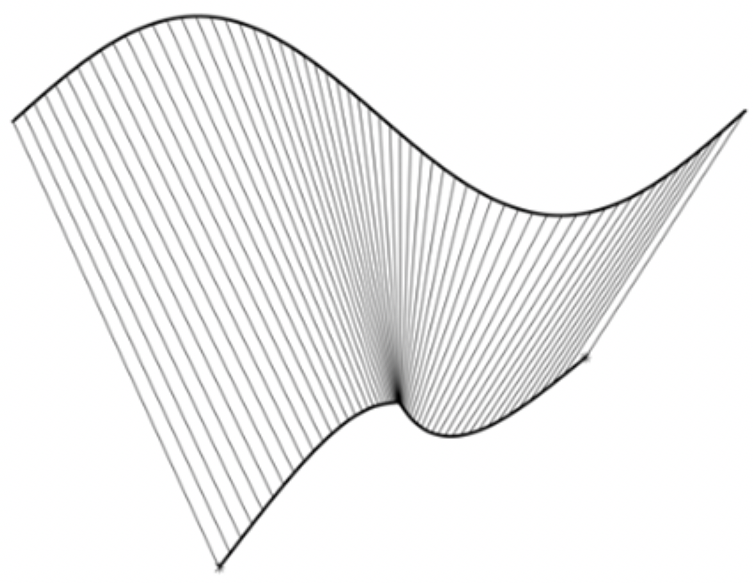}
\put(55,60){$C_0(t)$}
\put(20,1){$P_0$}
\put(80,25){$P_1$}
\end{overpic}\\
\centering (a)
\end{minipage}
\hfill
\begin{minipage}[b]{0.45\linewidth}
\includegraphics[width=0.71\textwidth]{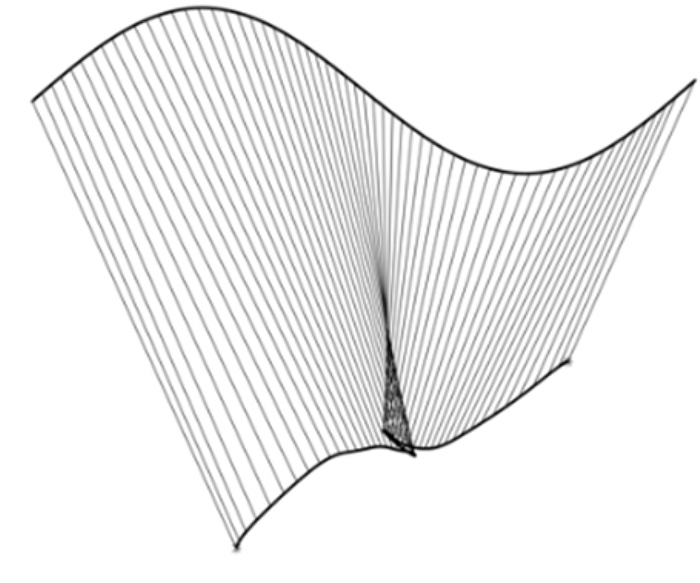}\\
\centering  (b)
\end{minipage}
}
\caption{The effect of the energy term $F_{Energy}$, $\lambda_{Width}=0$. (a) $\lambda_{Energy}=0.001$, $(\beta_{Max},\beta_{Ave})$=(1.82,0.51); (b) $\lambda_{Energy}=0$, $(\beta_{Max},\beta_{Ave})$=(5.24,0.48).}
\label{fig:fig2}
\end{figure}

\begin{figure}[t]
\centerline{
\scriptsize
\hfill
\begin{minipage}[b]{0.35\linewidth}
\centering
\begin{overpic}[width=1\textwidth]{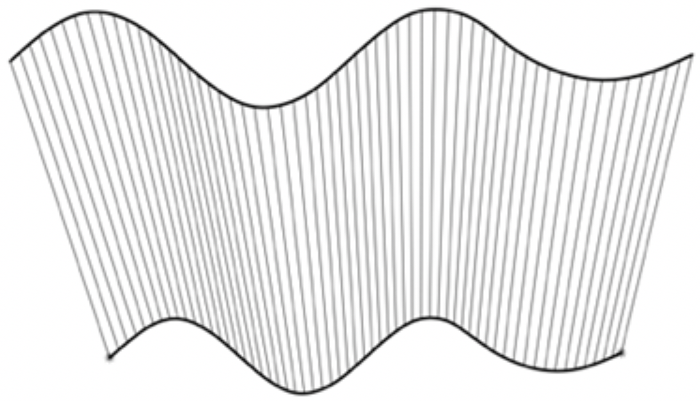}
\put(55,60){$C_0(t)$}
\put(7,5){$P_0$}
\put(90,5){$P_1$}
\end{overpic}\\
\centering (a)
\end{minipage}
\hfill
\begin{minipage}[b]{0.15\linewidth}
\includegraphics[width=1\textwidth]{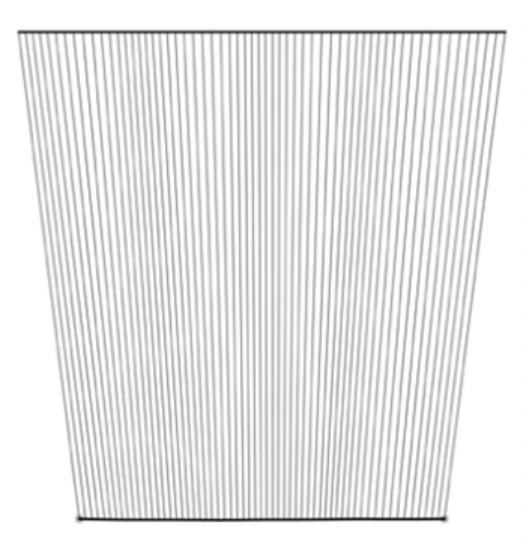}\\
\centering  (b)
\end{minipage}
\hfill
\begin{minipage}[b]{0.33\linewidth}
\centering
\begin{overpic}[width=1\textwidth]{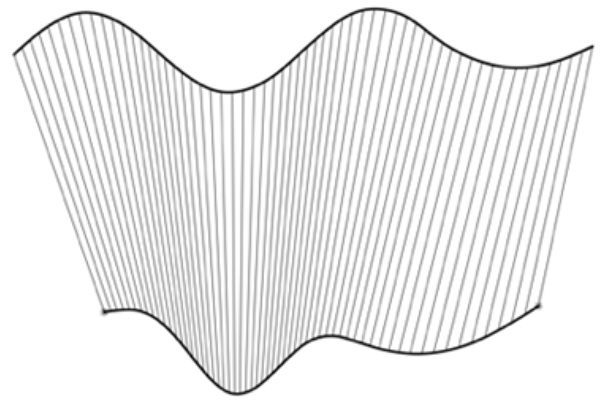}
\end{overpic}\\
\centering (c)
\end{minipage}
\hfill
\begin{minipage}[b]{0.13\linewidth}
\includegraphics[width=1\textwidth]{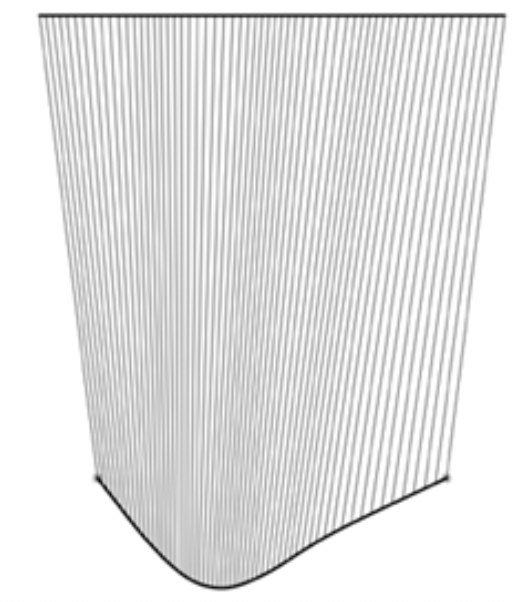}\\
\centering  (d)
\end{minipage}
}
\caption{The effect of the width variation control term $F_{Width}$, $\lambda_{Energy}=0.001$. (a) $\lambda_{Width}=0.00001$, $(\beta_{Max},\beta_{Ave})$=(2.88,0.65); (c) $\lambda_{Width}=0$, $(\beta_{Max}, \beta_{Ave})$=(4.63,0.73). (b),(d) are the top view of (a),(c) respectively.}
\label{fig:fig3}
\end{figure}

\emph{\textbf{Solution space regularization}}. The number of variables in optimization depends on the number of samplings on the surface. In fact, the number of samplings should be large enough to regularize the surface and it depends on the number of control points and the degree of the surface. In practice, to provide enough flexibility for maximizing surface developability, we use a large number of samplings (100).   Existing works have shown that when there are many variables in an optimization problem, the shape of the solution space is often not good, which makes the optimization process unstable and  even leads to an unsatisfactory solution in some cases. This problem can be solved by introducing a regularization term to control the shape of the solution space. Therefore, we introduce two shape-control terms as follows.

\begin{align}
F_{EnergyC_1} = \int_{0}^{1}\Vert C_1^{''}(t)\Vert^2 dt
\label{eq:eq3}
\end{align}
\begin{align}
F_{Width} = \sum_{j=0...M-1}(\Vert C_0(t_j)-C_1(t_j) \Vert^2-\Vert C_0(t_{j+1})-C_1(t_{j+1}) \Vert^2)^2
\label{eq:eq4}
\end{align}

\noindent where $F_{EnergyC_1}$ controls the smoothness of the curve  $C_1(t)$ and $F_{Width}$ controls the width variation  of the surface. Our experiments have shown that with these regularization terms the optimization is more stable and the shape of the resulting surface is superior to  those without regularization (refer to the experiments shown in Fig.~\ref{fig:fig2} and Fig.~\ref{fig:fig3}).

\emph{\textbf{Interior shape control of the surface}}. In order to control the interior shape of the surface, we introduce the following term, taking the effects of internal control rulings into account.

\begin{align}
F_{Interior} = \sum_{i=1...K-1}\Vert C_1(t_i)-P_i \Vert^2
\nonumber
\end{align}

\noindent where the parameter  $t_i$  is equal to the parameter  of the other endpoint of the control ruling line $L_i=\overline{C_0(t_i)P_i}$. This comes from Constraint 1 which requires two endpoints of one ruling have the same parameter. Finally, we arrive at the following objective function.

\begin{equation}
F_{1} = F_{Dev}+ \lambda_{Energy}F_{Energy}+\lambda_{Width}F_{Width}+\lambda_{Interior}F_{Interior}
\label{eqn:F1}
\end{equation}

By solving the minimization problem $F_1 \rightarrow min$, we can obtain a quasi-developable surface. The variables in the optimization are the independent normal vectors $N_k,k=0,...,M$ and the interior control points of curve $C_1(t)$ (the first and last control points remain fixed). This provides enough degrees of freedom, which enhances the ability to find a locally optimal quasi-developable surface. The algorithm is presented in Algorithm~\ref{alg:algorithm1}.

\begin{algorithm}[h] 
\caption{ Developable surface computation from a sequence of control rulings  and a fixed boundary curve.} 
\label{alg:algorithm1} 
\begin{algorithmic}[1] 
\Require
The sequence of control rulings $L_i=(Q_i,P_i),i=0,...,K$; 
The curve $C_0(t)$ which interpolate the data points $Q_i,i=0,...,K$.
The weights $\lambda_{Energy}$,$\lambda_{Width}$ and $\lambda_{Interior}$.
The sampling number for the normal vectors $M$.
\Ensure 
The interior control points of $C_1(t)$: $CP^1_{1},...,CP^1_{K-1}$. 
\State For each data point $P_i, i=0,...,K$, the parameter of the data point $Q_i$ on $C_0(t)$ is used for its parametrization.
\State A B-spline curve $C_1(t)$ satisfying $C_1(t_i)=P_i, i=0,...,K$ is computed by solving the linear system of equations.
\State A set of normal vectors $N_k$,$k=0,...,M$ are computed on the  B-spline surface $S$ bounded by $C_0(t)$ and $C_1(t)$, defined by Eq.(\ref{eq:eq1}).
\State Solve the minimization of Eq.(\ref{eqn:F1}) for the control points of $C_1(t)$ and the normal vectors $N_j,j=0,...,M$. \\
\Return $CP^1_{1},...,CP^1_{K-1}$.
\end{algorithmic} 
\end{algorithm}

With the output of the algorithm $CP^1_{1},...,CP^1_{K-1}$, together with the fixed end points $CP^1_{0}$ and $CP^1_{K}$, a B-spline surface defined by Eq.(\ref{eq:eq1}) is obtained. Note that the normal vectors $N_j$,$j=0,...,M$ are auxiliary variables used for simplifying the developability constraint equations and are not used for generating the resulting developable surface. In the following, we show some experimental results.

\emph{\textbf{Experiments (K=1)}}. In some applications, the first and the last ruling lines of the surface are specified as hard constraints, in addition to a fixed boundary curve $C_0(t)$. When there are no more control rulings other than the first and the last one, this case is characterized by $K=1$ in our definition. The first and last ruling line can be defined by two additional data points $P_0$ and $P_1$ in space and are denoted by $\overline{C_0(0)P_0}$ and $\overline{C_0(1)P_1}$. This design model is conceptually similar to the design of a Hermite curve by specifying the end data points and associated tangent vectors. In this case, the energy term $F_{EnergyC_1}$ and the width variation control term $F_{Width}$ have an important effect on controlling the interior  shape of the resulting surface which is illustrated by  the experimental results in Fig.~\ref{fig:fig2} and Fig.~\ref{fig:fig3}. The developability of the surface is evaluated by the warp angle between the normal vectors at the both ends of the ruling line. $\beta_{Max},\beta_{Ave}$  represent the maximum and average warp angle among a set of sampling rulings, respectively.

Observing the results  in Fig.~\ref{fig:fig2}, we find that the energy term $F_{Energy}$ can efficiently control the shape of the boundary curve and avoid self-overlapping of the surface, without sacrificing the developability of the surface. In fact, the resulting surface achieves a higher degree of developability  than  the result without the energy term.   This is because the energy term $F_{Energy}$ leads to a finer shape of the solution space and consequentially a stable optimization process which avoids unsatisfactory local minimums. Observing the results shown in Fig.~\ref{fig:fig3}, the width variation control term $F_{Width}$  makes the resulting surface have a smooth variation of the width of ruling lines with improved surface developability, owing to the same reason as the energy term $F_{Energy}$.

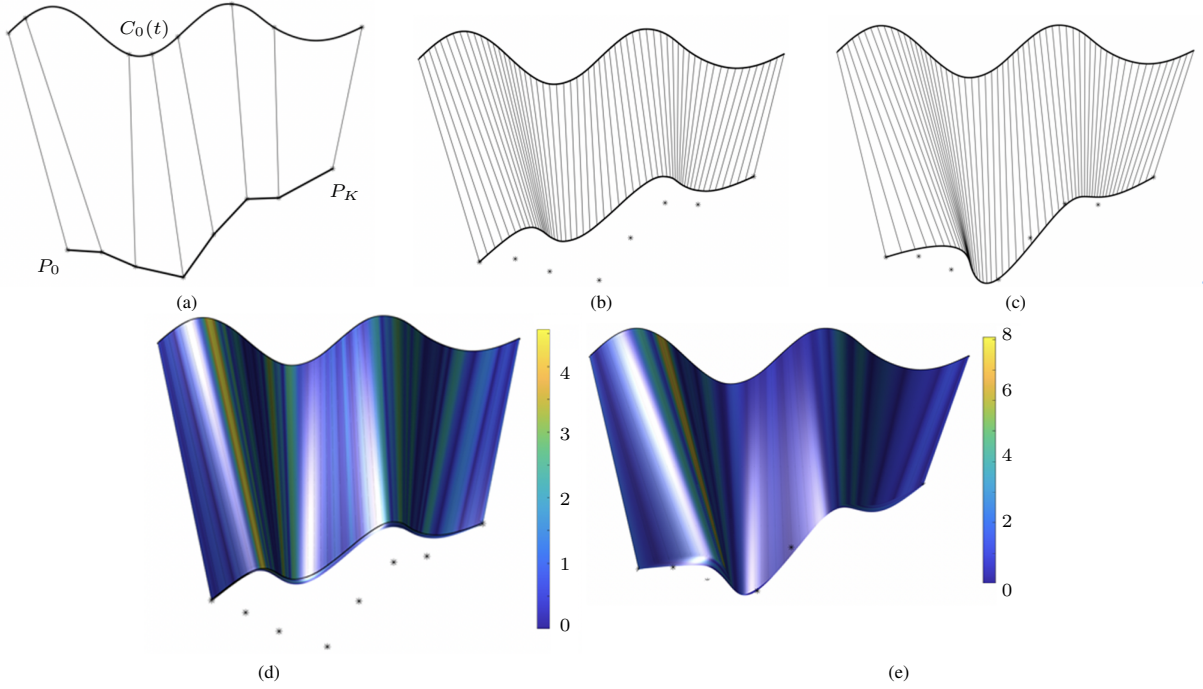
\begin{figure}[t]
\centerline{
\scriptsize
\hfill
\begin{minipage}[b]{0.3\linewidth}
\centering
\begin{overpic}[width=1\textwidth]{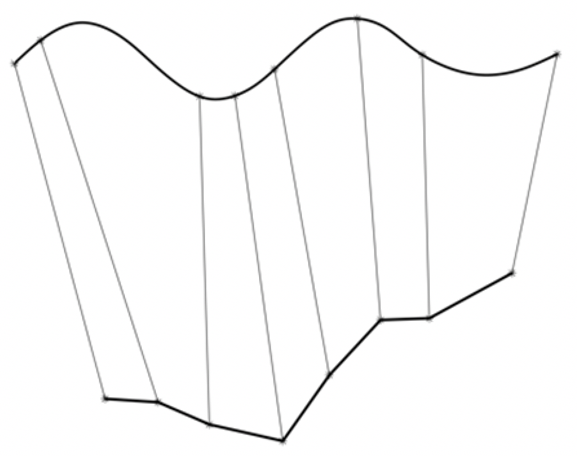}
\put(32,68){$C_0(t)$}
\put(10,4){$P_0$}
\put(88,24){$P_K$}
\end{overpic}\\
\centering (a)
\end{minipage}
\hfill
\begin{minipage}[b]{0.3\linewidth}
\includegraphics[width=1\textwidth]{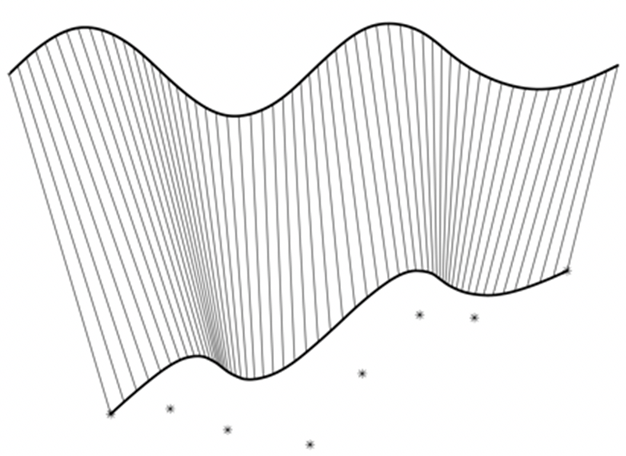}\\
\centering  (b)
\end{minipage}
\hfill
\begin{minipage}[b]{0.3\linewidth}
\includegraphics[width=1\textwidth]{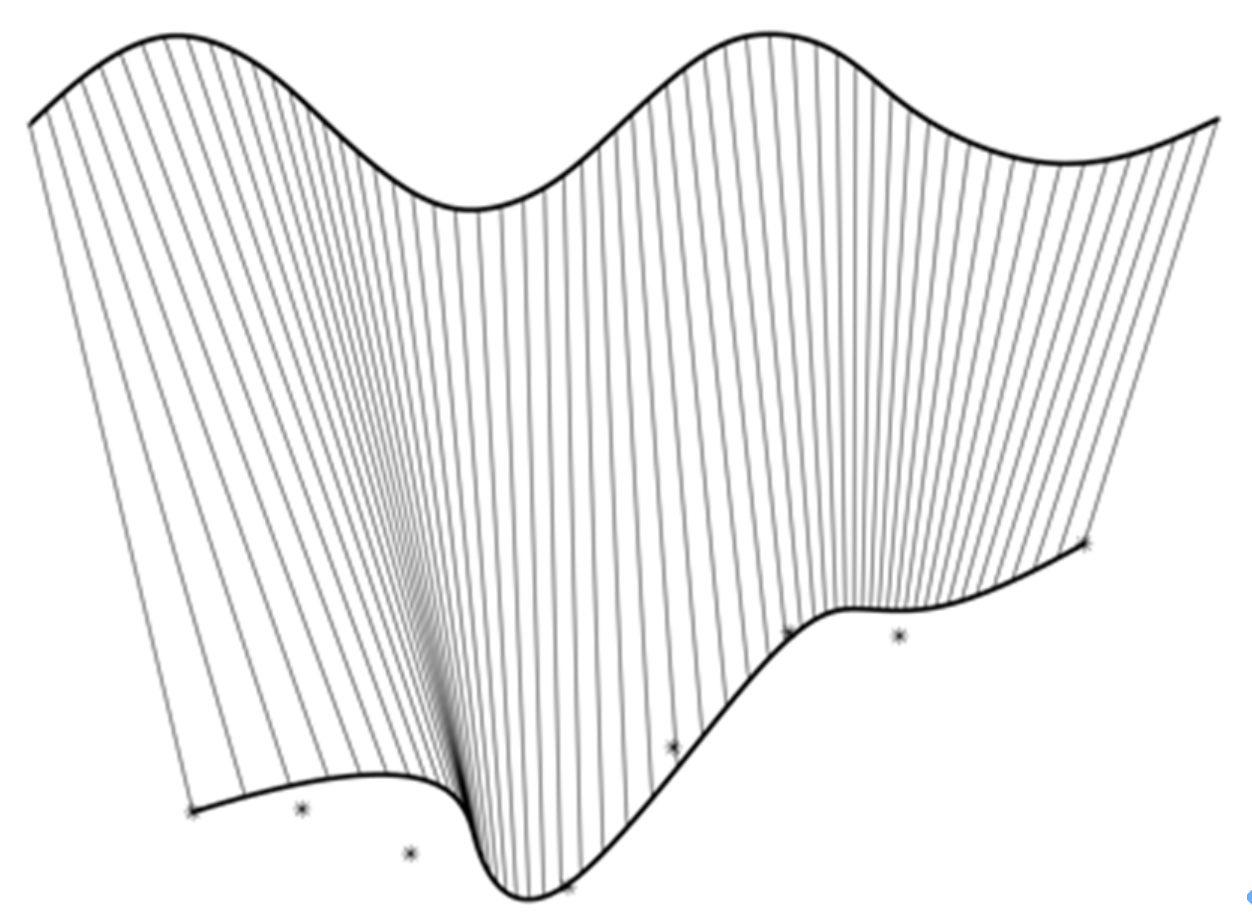}\\
\centering  (c)
\end{minipage}
}
\centerline{
\scriptsize
\hfill
\begin{minipage}[b]{0.5\linewidth}
\flushright
\begin{overpic}[width=0.7\textwidth]{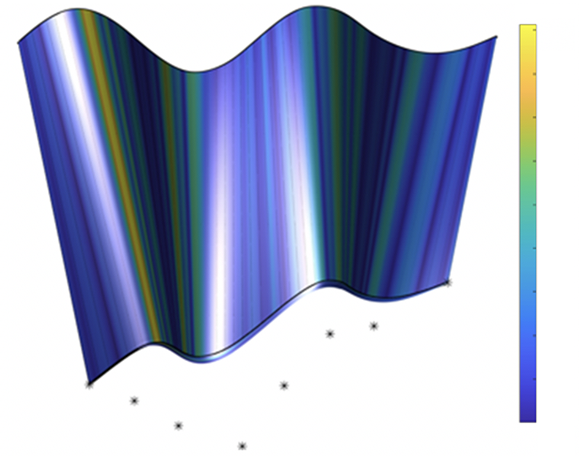}
\put(95,6){$0$}
\put(95,20){$1$}
\put(95,35){$2$}
\put(95,50){$3$}
\put(95,64){$4$}
\end{overpic}\\
\centering (d)
\end{minipage}
\hfill
\begin{minipage}[b]{0.5\linewidth}
\flushleft 
\begin{overpic}[width=0.7\textwidth]{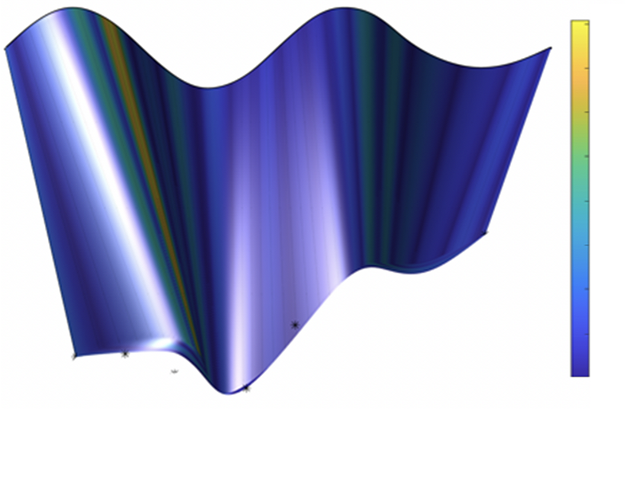}
\put(95,14){$0$}
\put(95,30){$2$}
\put(95,45){$4$}
\put(95,60){$6$}
\put(95,73){$8$}
\end{overpic}\\
\centering (e)
\end{minipage}
}
\caption{The effect of $F_{Interior}$, $\lambda_{Energy}=0.001,\lambda_{Width}=0.00001$.(a) Input curve and given rulings. (b)(d) Result without the constraint of interior rulings. $\lambda_{Interior}=0$, $(\beta_{Max},\beta_{Ave})$=(4.14,1.12), $(D_{Max},D_{Ave})$=(0.198,0.099); (c)(e) Result with the constraint of interior rulings. $\lambda_{Interior}=1$, $(\beta_{Max},\beta_{Ave})$=(7.94,1.37), $(D_{Max},D_{Ave})$=(0.085,0.032).}
\label{fig:fig4}
\end{figure}

\emph{\textbf{Experiments (K\textgreater1)}}. When $K\textgreater1$, the interior rulings other than the end ones can be used to control the interior shape of the resulting surface. The experiments in Fig.~\ref{fig:fig4} show the difference between the results with  and without the constraint of interior rulings. The upper boundary curve  $C_0(t)$ in Fig.~\ref{fig:fig4}(a) is the given curve which is fixed. Fig.~\ref{fig:fig4}(b) and Fig.~\ref{fig:fig4}(c) show the developable surfaces without and with the constraint of interior rulings, respectively. The color coding in Fig.~\ref{fig:fig4} indicates the magnitude of the warp angle. We also measure the distances from the endpoints of rulings to the final boundary curve. $D_{Max},D_{Ave}$  represent the maximum and average distance, respectively. We conclude from the experimental results that the resulting surface can be controlled by interior rulings without giving up much developability. However, the constraint of interior rulings indeed has negative effect on surface developability. This means the hard interpolations to the  boundary curve $C_0(t)$ and two terminal  rulings greatly restrict the solution space of developable surfaces.  In Sec.~\ref{sec:relaxed2boundary} we will discuss the method relaxing the constraints of boundary conditions.

\subsection{\textbf{Developable surface design from control rulings with two relaxed boundary curves}}
\label{sec:relaxed2boundary}

\begin{figure}[ht]
\centerline{
\scriptsize
\hfill
\begin{minipage}[b]{0.32\linewidth}
\centering
\begin{overpic}[width=1\textwidth]{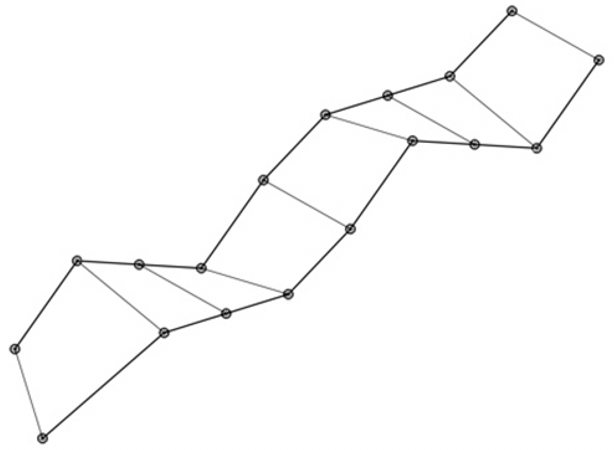}
\end{overpic}\\
\centering (a)
\end{minipage}
\hfill
\begin{minipage}[b]{0.3\linewidth}
\includegraphics[width=1\textwidth]{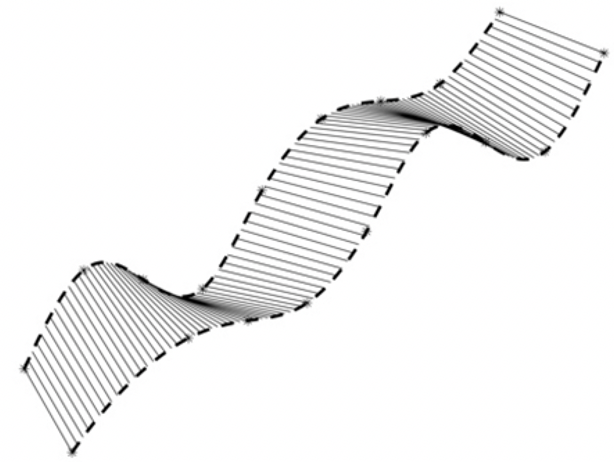}\\
\centering  (b)
\end{minipage}
\hfill
\begin{minipage}[b]{0.3\linewidth}
\includegraphics[width=1\textwidth]{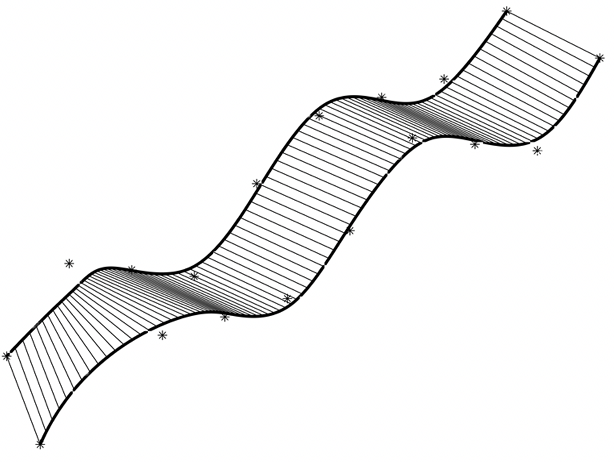}\\
\centering  (c)
\end{minipage}
}
\centerline{
\scriptsize
\hfill
\begin{minipage}[b]{0.3\linewidth}
\centering
\begin{overpic}[width=1\textwidth]{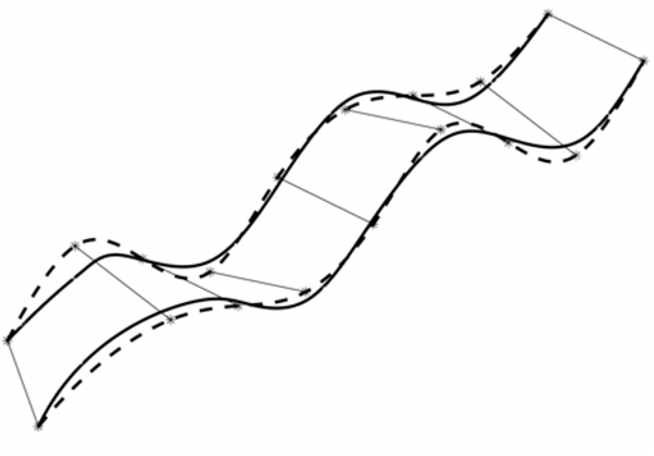}
\end{overpic}\\
\centering (d)
\end{minipage}
\hfill
\begin{minipage}[b]{0.26\linewidth}
\centering
\begin{overpic}[width=1\textwidth]{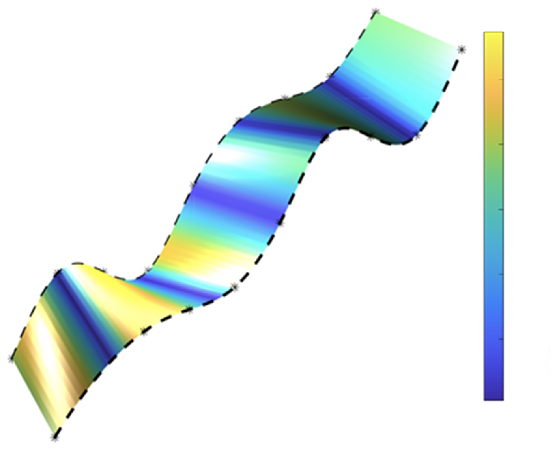}
\put(92,7){$0$}
\put(92,29){$10$}
\put(92,54){$20$}
\put(92,73){$30$}
\end{overpic}\\
\centering (e)
\end{minipage}
\hfill
\begin{minipage}[b]{0.3\linewidth}
\centering
\begin{overpic}[width=1\textwidth]{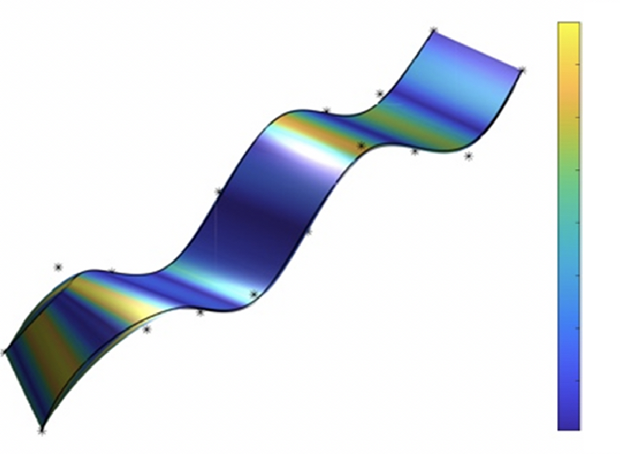}
\put(95,3){$0$}
\put(95,20){$1$}
\put(95,38){$2$}
\put(95,55){$3$}
\put(95,70){$4$}
\end{overpic}\\
\centering (f)
\end{minipage}
}
\centerline{
\scriptsize
\hfill
\begin{minipage}[b]{0.4\linewidth}
\flushright
\begin{overpic}[width=0.9\textwidth]{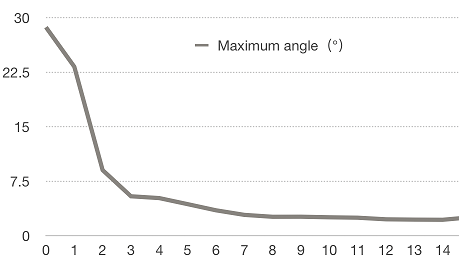}
\end{overpic}\\
\centering (g)
\end{minipage}
\hfill
\begin{minipage}[b]{0.4\linewidth}
\flushleft
\begin{overpic}[width=0.88\textwidth]{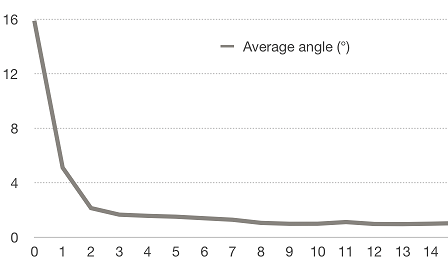}
\end{overpic}\\
\centering (h)
\end{minipage}
\hfill
}
\caption{Developable surface from control rulings with both  curves  relaxed. $\lambda_{Energy}=0.00001, \lambda_{Width}=0.00001, \lambda_{Intepolate}=1.$ (a) Input control  rulings forming a quad strip. (b)(e) Initial interpolation surface, $(\beta_{Max},\beta_{Ave})$=(28.68,15.88); (c)(f) Resulting surface. $(\beta_{Max}, \beta_{Ave})$=(3.91,1.27), $(D_{Max},D_{Ave})$=(0.025,0.0089). (d) The initial boundary curves (dotted curves) and the result boundary curves (solid curves) are shown together for comparison. The convergence behaviour of the algorithm is shown in (g) and (h). (g) Maximum warp angle vs. iteration. (h) Average warp angle vs. iteration.}
\label{fig:fig5}
\end{figure}

In many applications (such as ship-hull design), the input curves are not required to be the exact boundary curves of the resulting surface. A distance tolerance is often allowed whose value depends on specific applications.  Therefore, a method which can find a surface with a high degree of developability  with its boundary curves perturbed from the original design curves are more practically useful.   We devise  such a method in which both boundary curves are  variables  in optimization and solve for perturbed boundary curves to define a developable surface.  The problem is described as follows.

Given a set of ordered rulings $L_i=(Q_i,P_i)(i=0,...,K)$, $Q_i,P_i$ are the given data points. The objective is to construct the curves $C_0(t),C_1(t)$ fitting the data points $Q_i,P_i,(i=0,...,K)$, respectively, satisfying $C_0(0)=Q_0,C_1(0)=P_0,C_0(1)=Q_K,C_1(1)=P_K$ so that the surface $S$ (Eq.(\ref{eq:eq1})) bounded by $C_0(t)$ and $C_1(t)$ achieves a high degree of developability.

Viewing the two curves $C_0(t)$ and $C_1(t)$ as variables in optimization provides more degrees of freedom and increases the possibility of finding a surface with a higher degree of developability. As we have explained, this  problem of fitting developable surfaces to a set of ruling lines is equivalent to  fitting two curves to the data points $Q_i$ and $P_i $, $i=0,...,K$ with certain constraints.  It is well-known that the parametrization of the data points is a key issue in curve fitting. In our specific problem, data parametrization is even more tricky than in ordinary curve fitting because the data points $Q_i$ and $P_i$ are required to have the same parameter,  which is a requirement from Constraint 1. Therefore, the key issue is how to find the optimal parameter value $t_i$ for $Q_i$ and $P_i$.  We propose an  algorithm which solves for  $C_0(t)$ and $C_1(t)$  and the parameters  $t_i$ through  an iterative process. The main steps of this algorithm are described as follows.

\begin{enumerate}
\item Step1. Parametrize the rulings to make the data points $Q_i, i=0,...,K$ and $P_i, i = 0, ..., K$ have the same parameter values $t_0, ..., t_K$, respectively. We employ the centripedal parametrization  to the data points $Q_i, i=0,...,K$ and $P_i, i = 0, ..., K$, and   obtain the parameter values  $u_0,...,u_K$ and $v_0,...,v_K$, respectively.  Then the  average parameter value   of the  two endpoints of a  ruling is used for both endpoints. Precisely,  $t_i=\frac{1}{2}(u_i+v_i)$ is used as the parameter value for both $Q_i$ and $P_i$.

\item Step2. Construct the interpolation B-spline curves $\widetilde{C_0}(t)$ and $\widetilde{C_1}(t)$ satisfying $\widetilde{C_0}(t_j)=Q_j$ and $\widetilde{C_1}(t_j)=P_j, j=0,...,K$. Generally, the surface $S$ bounded by $\widetilde{C_0}(t)$ and $\widetilde{C_1}(t)$ as defined by Eq.(\ref{eq:eq1})  is not a developable surface. Therefore, we need to to improve the developability of the surface by relaxing the constraint of curve interpolation.  We define the following closeness term.

\begin{equation}
F_{Closeness} = \sum_{i=1...K-1}(\Vert C_1(t_i)-P_i \Vert^2+\Vert C_0(t_{i})-Q_i) \Vert)^2
\label{eq:close}
\end{equation}

Note that the end control points of both curves are fixed and are not treated as variables in optimization, to confirm hard interpolation to the specified terminal rulings. Then we define the following objective function.

\begin{align}
F_{2} = F_{Dev}+ \lambda_{Closeness}F_{Closeness}+\lambda_{Energy}(F_{EnergyC_0}+F_{EnergyC_1})+\lambda_{Width}F_{Width}
\label{eq:F2}
\end{align}

\noindent where $F_{EnergyC_0}$ is an energy term for $C_0(t)$ whose definition is samilar to the definition of $F_{EnergyC_1}$ (Eq.(\ref{eq:eq3})). For the definition of $F_{Width}$, refer to Eq.(\ref{eq:eq4}).

\item Step3.  Solve the minimization problem $F_2 \rightarrow min$, we can obtain two updated curves $\widetilde{C_0}(t)$ and $\widetilde{C_1}(t)$ (the same symbols are used for the curves before and after update). 

In order to provide a termination criterion of the iterative process, we evaluate the developability of the surface $S$  bounded by the updated curves $\widetilde{C_0}(t)$ and $\widetilde{C_1}(t)$. If the developability requirements are met, or the improvement of  developability  by this iteration is ignorable,  the algorithm can be stopped; Otherwise goto Step3.

\item Step4. Update the parameter values of  the data points. For the data point $Q_j$,  a parameter value $u_j$ is obtained which is the parameter of the nearest point (foot point) on $\widetilde{C_0}(t)$. Similarly, for the data point $P_j$, the parameter value $v_j$ of the nearest point (foot point) on $\widetilde{C_1}(t)$ is computed. For foot point computation, a Newton-like iteration method is used~\cite{Hoschek1993}. Then the value $t_j=\frac{1}{2}(u_j+v_j)$ is used as the updated parameter for  both $Q_j$ and $P_j$. Goto Step3.
\end{enumerate}

\begin{algorithm}[h] 
\caption{Developable surface computation from a sequence of control rulings.} 
\label{alg:algorithm2} 
\begin{algorithmic}[1] 
\Require
The sequence of control rulings $L_i=(Q_i,P_i),i=0,...,K$; 
The weights $\lambda_{Closeness}$,$\lambda_{Energy}$ and $\lambda_{Width}$.
The sampling number for the normal vectors $M$.
\Ensure 
The interior control points of $C_0(t)$: $CP^0_{1},...,CP^0_{K-1}$. The interior control points of $C_1(t)$: $CP^1_{1},...,CP^1_{K-1}$. 
\State Compute the parametrization for the data points $Q_i, i=0,...,K$, and $P_i,i=0,...,M$, which are $t_i$,$i=0,...,K$.
\State Two B-spline curves are computed as the initialization by solving the linear systems of equations  $C_0(t_i)=Q_i, i=0,...,K$ and  $C_1(t_i)=P_i, i=0,...,K$, respectively.
\Repeat
\State Solve the minimization of Eq.(\ref{eq:F2}) for the control points of $C_0(t)$ and $C_1(t)$, and the normal vectors $N_j,j=0,...,M$. 
\State Compute the parameters $t_i$, $i=0,...,K$ of the data points $Q_i$ and $P_i$, $i=0,...,K$ by projecting the data points to the corresponding curves $C_0(t)$ and $C_1(t)$, respectively.
\Until The improvement of the surface is smaller than a predefined tolerance.\\
\Return $CP^0_{1},...,CP^0_{K-1}$ and $CP^1_{1},...,CP^1_{K-1}$.
\end{algorithmic} 
\end{algorithm}

The algorithm is summarized in Algorithm~\ref{alg:algorithm2}. The output of the algorithm combined with the fixed end  points give us the control points of $C_0(t)$ and $C_1(t)$, i.e.,  $CP^0_{0}$,...,$CP^0_{K}$  and  $CP^1_{0}$,...,$CP^1_{K}$.  The surface $S$ is thus obtained by the definition of Eq.(\ref{eq:eq1}), bounded by $C_0(t)$ and $C_1(t)$.


An experiment is shown in Fig.~\ref{fig:fig5}. Given a set of rulings (Fig.~\ref{fig:fig5}(a)), we first construct two initial interpolation B-spline curves $\widetilde{C_0}(t)$, $\widetilde{C_1}(t)$ and the initial interpolation surface $S$ (Fig.~\ref{fig:fig5}(b)). Fig.~\ref{fig:fig5}(c) shows the result boundary curves and the resulting surface. Fig.~\ref{fig:fig5}(e) and Fig.~\ref{fig:fig5}(f) shows the color coding of warp angles of the surfaces in Fig.~\ref{fig:fig5}(b) and Fig.~\ref{fig:fig5}(c), respectively.  Within the range of unit 1, the maximum distance from given data points to the resulting curves $C_0(t)$ and $C_1(t)$ is 0.025, and the average distance is 0.0089, indicating that the surface achieves high degrees of both developability and approximation to input ruling lines. Fig.~\ref{fig:fig5}(d) shows the difference between initial interpolation curves (dotted curves) and result boundary curves (solid curves). The convergence behaviour of our algorithm is shown in Fig.~\ref{fig:fig5}(g)(h). It can be seen that the algorithm in Sec.\ref{sec:relaxed2boundary} has high convergence rate and high stability.

\subsection{\textbf{Designing the control rulings}}
\label{sec:controlRulingDesign}
The algorithms proposed in this work depend on a sequence of line segments  serving as  the control rulings of  the surface. Basically,  the lines can be specified in space freely  by the designer. However, with arbitrary line segments in space, the distance between the control lines and the  resulting developable surface may be large which  is not desired in an interactive design process.  

To reduce the difference between the original design rulings and the resulting surface, while still achieving a high degree of developability,  additional constraints in the design of input rulings should be considered.  It has been shown that planar quadrilateral strips can be viewed as the discretization  of smooth developable surfaces~\cite{Liu2006Geometric}. Therefore, in the design process of the control rulings, we require that every pair of adjacent rulings be coplanar.  Precisely, if the given rulings are denoted by $L_i=(P_i,Q_i), i=0,...,K$, we require the quadrilaterals $P_iQ_iQ_{i+1}P_{i+1}$ bounded by two adjacent rulings $L_i$,$L_{i+1}$ to be coplanar. 

\begin{figure}[t]
\centerline{
\scriptsize
\hfill
\begin{minipage}[b]{0.3\linewidth}
\centering
\begin{overpic}[width=1\textwidth]{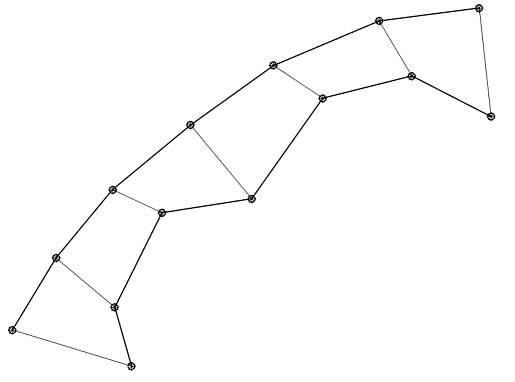}
\end{overpic}\\
\centering (a)
\end{minipage}
\hfill
\begin{minipage}[b]{0.3\linewidth}
\includegraphics[width=1\textwidth]{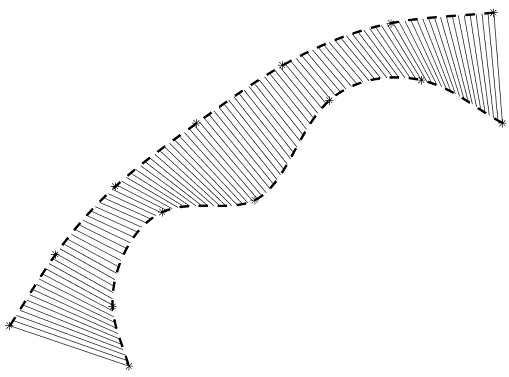}\\
\centering  (b)
\end{minipage}
\hfill
\begin{minipage}[b]{0.3\linewidth}
\includegraphics[width=1\textwidth]{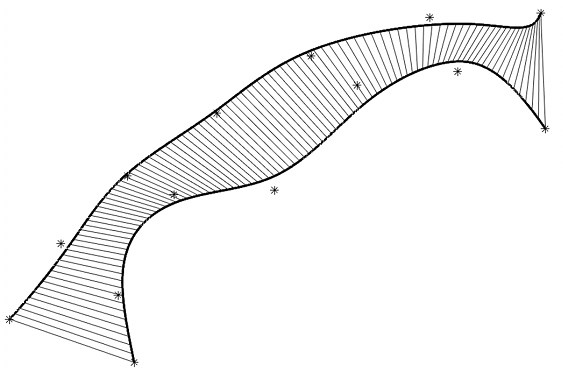}\\
\centering  (c)
\end{minipage}
}
\centerline{
\scriptsize
\hfill
\begin{minipage}[b]{0.3\linewidth}
\centering
\begin{overpic}[width=1\textwidth]{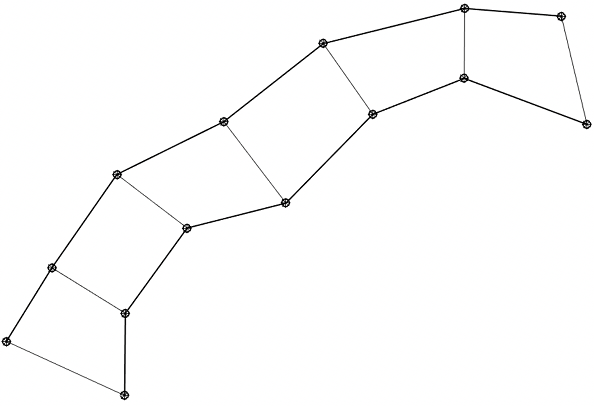}
\end{overpic}\\
\centering (d)
\end{minipage}
\hfill
\begin{minipage}[b]{0.3\linewidth}
\includegraphics[width=1\textwidth]{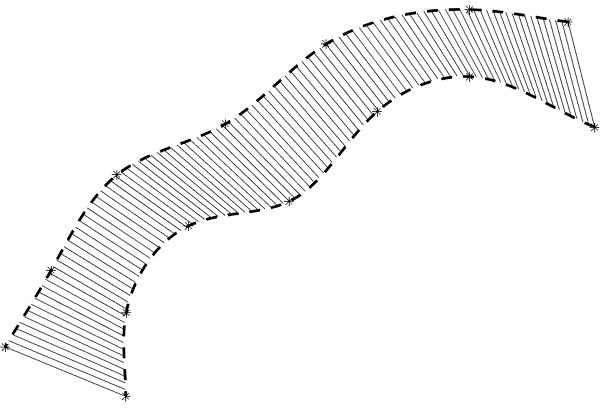}\\
\centering  (e)
\end{minipage}
\hfill
\begin{minipage}[b]{0.3\linewidth}
\includegraphics[width=1\textwidth]{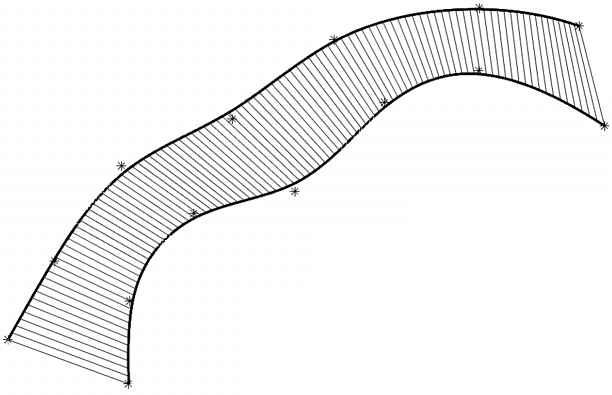}\\
\centering  (f)
\end{minipage}
}
\caption{The first row shows the experiments from  control rulings without adjacent coplanarity. $\lambda_{Energy}$=0.00001, $\lambda_{Width}$=0.00001, $\lambda_{Closeness}$=0.1. (a) The designed control rulings. (b) Initial interpolation surface $(\beta_{Max},\beta_{Ave})$=(41.10,25.26); (c) Resulting surface $(\beta_{Max},\beta_{Ave})$=(8.98,1.67), $(D_{Max},D_{Ave})$=(0.045,0.022).  The second row shows the experiment from control rulings with adjacent coplanarity. $\lambda_{Energy}$=0.00001, $\lambda_{Width}$=0.00001, $\lambda_{Closeness}$=0.1. (d) The designed control rulings. (e) Initial interpolation surface $(\beta_{Max},\beta_{Ave})$=(9.07,2.51); (d) Resulting surface $(\beta_{Max},\beta_{Ave})$=(0.75,0.19), $(D_{Max},D_{Ave})$=(0.014,0.006).}
\label{fig:fig9}
\end{figure}

In Fig.~\ref{fig:fig9}(a)-(c), the  control rulings are designed arbitrarily and we observe that  it is not always guaranteed that  a resulting surface with a high  degree of developability can meet the original design intention tightly. That is, a compromise between the original design intention of shape and a high degree of developability should be made.

Fig.~\ref{fig:fig9}(d)-(f) give the experimental results with the original control rulings similar to the one in Fig.~\ref{fig:fig9}(a), which however satisfy the coplanarity constraint. From the experimental results in Fig.~\ref{fig:fig9}(d)-(f), we observe that 
 we can obtain a developable surface better meeting the original design intention with input lines forming a planar quadrilateral strip. The developability of the resulting surface based on coplanar lines is about an order of magnitude better than the results of non-planar lines with even better approximation quality to the control rulings.

From these experiments, we can suggest the following procedural way of control ruling design.  Once the $i$th control ruling $L_i$ is ready, the $(i+1)$th ruling should be defined on the plane containing $L_i$. Note that although this process depends on a sequence of planes, this  method is essentially different to the existing methods working with planes tangent to the developable surface in that our method works directly in the design space which has more geometric intuition than working in the dual space. This algorithm of control ruling design is presented in Algorithm~\ref{alg:algorithm3}.

\begin{algorithm}[h] 
\caption{The design process of the sequence of control rulings.} 
\label{alg:algorithm3} 
\begin{algorithmic}[1] 
\Require
A target shape in the mind of the designer.
\Ensure 
The sequence of straight line segments in space $L_i=(Q_i,P_i)$, $i=0,...,K$.
\State The first line segment $L_0$ is defined by interactively specifying two data points $Q_0$ and $P_0$ in space. 
\State i=0;
\Repeat
\State A plane $\Omega_i$ containing $L_i$ is defined by specifying a line segment $W_i=(A_i, B_i)$ in space  where $A_i$ is  a data point on $L_i$.
\State The $(i+1)$th line segment is defined by specifying two data points on the plane $\Omega_i$. 
\State i++;
\Until The current  line segment is specified as the last line segment. \\
\Return  $L_i=(Q_i,P_i)$, $i=0,...,K$.
\end{algorithmic} 
\end{algorithm}

\section{Method evaluation and discussions}

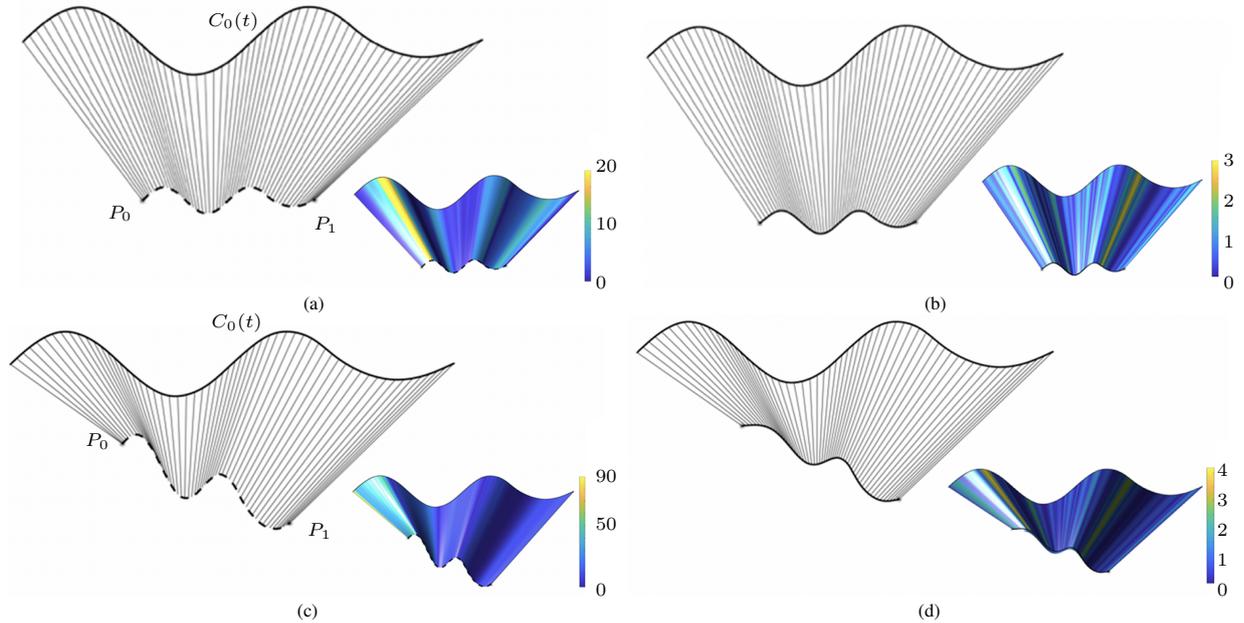
\begin{figure}[t]
\centerline{
\scriptsize
\hfill
\begin{minipage}[b]{0.47\linewidth}
\centering
\begin{overpic}[width=1\textwidth]{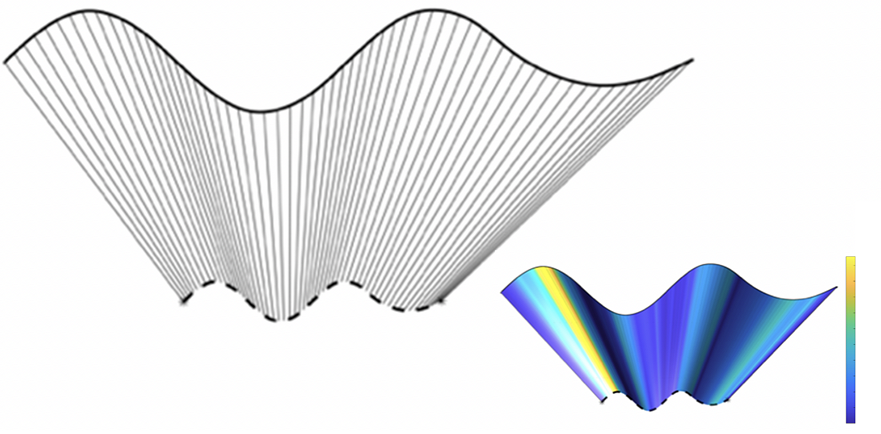}
\put(32,45){$C_0(t)$}
\put(15,12){$P_0$}
\put(50,10){$P_1$}
\put(98, 0){$0$}
\put(98, 10){$10$}
\put(98, 20){$20$}
\end{overpic}\\
\centering (a)
\end{minipage}
\hfill
\begin{minipage}[b]{0.47\linewidth}
\begin{overpic}[width=1\textwidth]{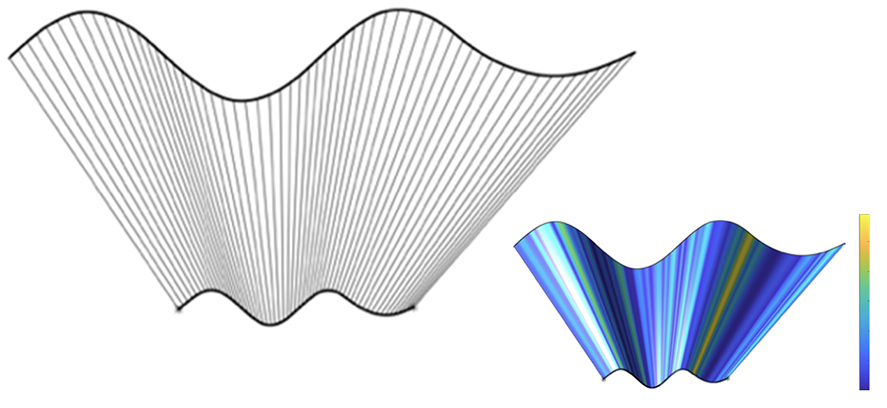}
\put(99, 0){$0$}
\put(99, 7){$1$}
\put(99, 14){$2$}
\put(99, 21){$3$}
\end{overpic}\\
\centering  (b)
\end{minipage}
}
\centerline{
\scriptsize
\hfill
\begin{minipage}[b]{0.48\linewidth}
\centering
\begin{overpic}[width=1\textwidth]{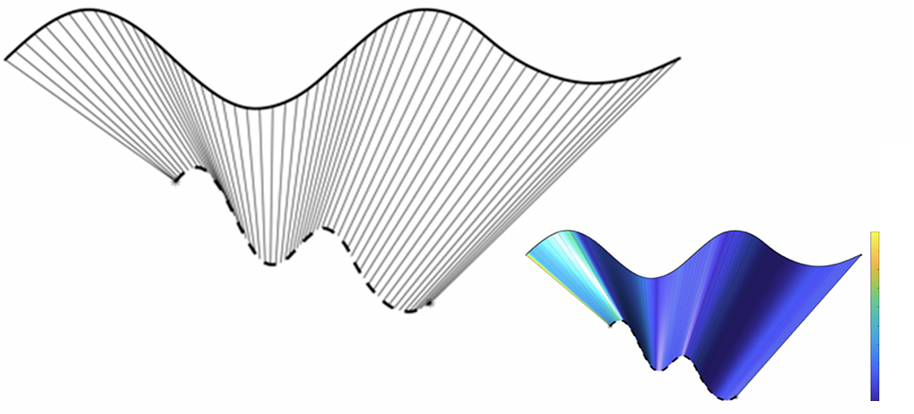}
\put(34,45){$C_0(t)$}
\put(13,25){$P_0$}
\put(50,10){$P_1$}
\put(98,0){$0$}
\put(98,11){$50$}
\put(98,19){$90$}
\end{overpic}\\
\centering (c)
\end{minipage}
\hfill
\begin{minipage}[b]{0.48\linewidth}
\begin{overpic}[width=1\textwidth]{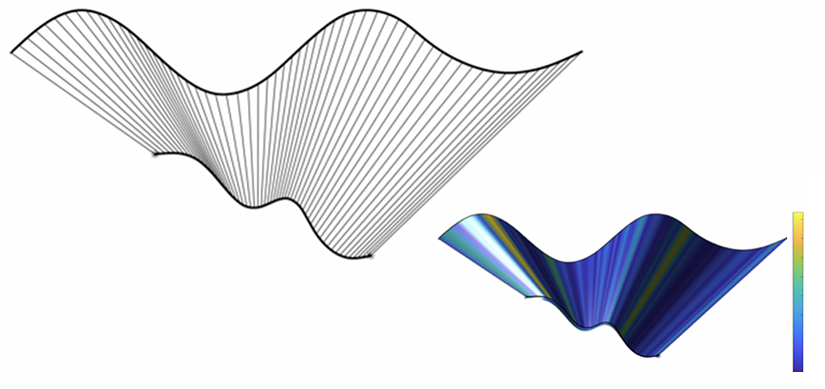}
\put(98,0){$0$}
\put(98,5){$1$}
\put(98,10){$2$}
\put(98,15){$3$}
\put(98,20){$4$}
\end{overpic}\\
\centering  (d)
\end{minipage}
}
\caption{The first row shows the experiment with one fixed boundary curve. $\lambda_{Energy}$=0.001, $\lambda_{Width}$=0.00001. (a) Initial surface $(\beta_{Max},\beta_{Ave})$=(20.63,4.86); (b) Resulting surface $(\beta_{Max},\beta_{Ave})$=(2.88,0.70). The second row shows the experiment with one fixed boundary curve. $\lambda_{Energy}$=0.001, $\lambda_{Width}$=0.00001. (a) Initial surface $(\beta_{Max},\beta_{Ave})$=(90.00,16.37); (b) Resulting surface $(\beta_{Max},\beta_{Ave})$=(3.95,0.89). The color coding of each  surface is based on the range of its own maximum and minimum warp angle.}
\label{fig:fig7}
\end{figure}

\subsection{The flexibility of our method in choosing the terminal ruling lines}

Fig.~\ref{fig:fig7} shows the experiments with one fixed boundary curve, using the algorithm in Sec.~\ref{sec:onefixboundary} to construct a developable surface. The inputs include two data points ${P_0,P_1}$ and one fixed boundary curve $C_0(t)$ which is the same as the fixed curve in Fig.~\ref{fig:fig3}.  The goal is to construct a B-spline boundary curve $C_1(t)$ whose endpoints are ${P_0,P_1}$ and the  surface $S$  bounded by $C_0(t)$ and $C_1(t)$ as defined by Eq.(\ref{eq:eq1}) achieves a high degree of developability. The boundary curve $C_1(t)$ are initialized  by transforming the control points of $C_0(t)$ by rotation, translation and scaling operations such that the end points of the transformed curve coincide with $P_0$ and $P_1$.  Fig.~\ref{fig:fig4}(b), Fig.~\ref{fig:fig7} show surfaces of large  developability with the same fixed boundary curve $C_0(t)$ and various positions of $P_0$ and $P_1$.  The results demonstrate that by fixing one boundary curve and two terminal rulings, there are still enough degrees of freedom for generating  a high degree of developability.

When more flexible control over the interior shape of the surface is desired, the interior rulings should be taken into consideration. This  is characterized in our definition by $K>1$ and $\lambda_{Closeness} > 0$. The experiment in Fig.~\ref{fig:fig4} has shown that when interior control rulings are considered, fixing one boundary curve does not provide enough flexibility.   In this case, the boundary curve interpolation constraints should be relaxed to provide a high possibility to find a surface with a high degree of developability, which is realized using the algorithm in Sec.~\ref{sec:relaxed2boundary}.


\subsection{The superiority of relaxing both boundary curves}
\begin{figure}[htb]
\centerline{
\scriptsize
\hfill
\begin{minipage}[b]{0.3\linewidth}
\centering
\begin{overpic}[width=1\textwidth]{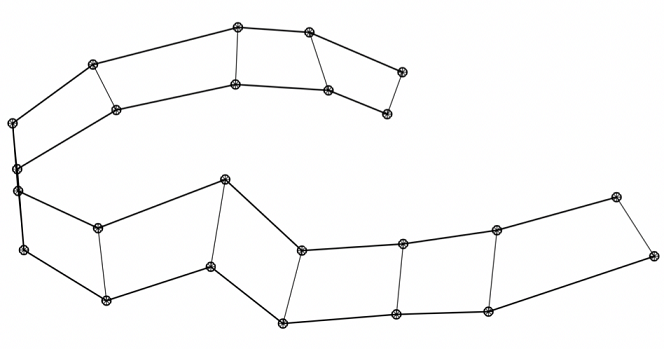}
\end{overpic}\\
\centering (a)
\end{minipage}
\hfill
\begin{minipage}[b]{0.3\linewidth}
\includegraphics[width=1\textwidth]{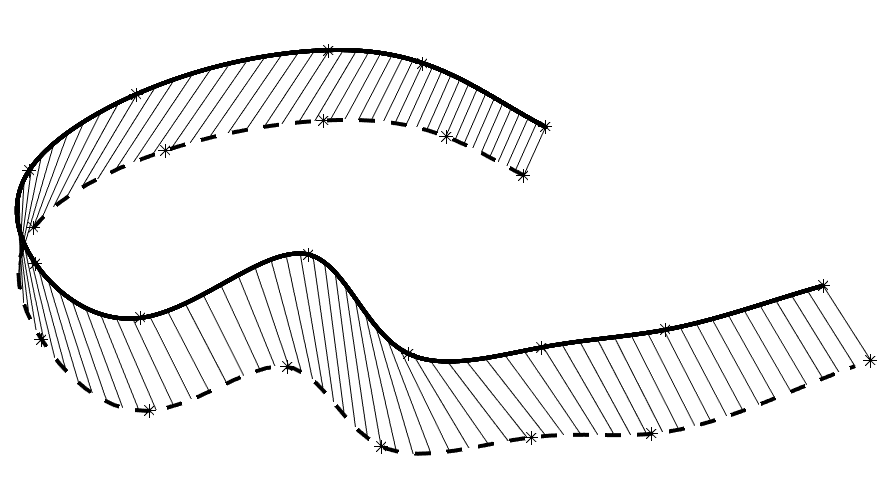}\\
\centering  (b)
\end{minipage}
\scriptsize
\hfill
\begin{minipage}[b]{0.3\linewidth}
\centering
\begin{overpic}[width=1\textwidth]{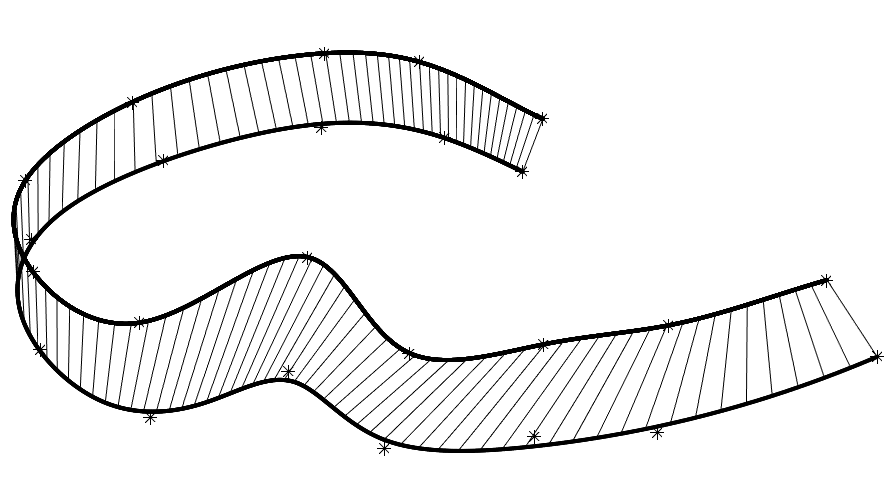}
\end{overpic}\\
\centering (c)
\end{minipage}
\hfill
}
\centerline{
\scriptsize
\begin{minipage}[b]{0.3\linewidth}
\includegraphics[width=1\textwidth]{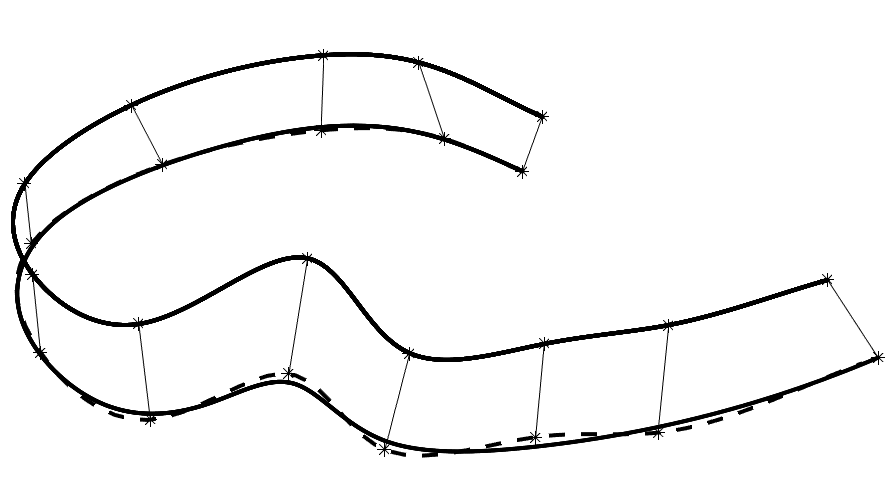}\\
\centering  (d)
\end{minipage}
\hfill
\begin{minipage}[b]{0.3\linewidth}
\centering
\begin{overpic}[width=1\textwidth]{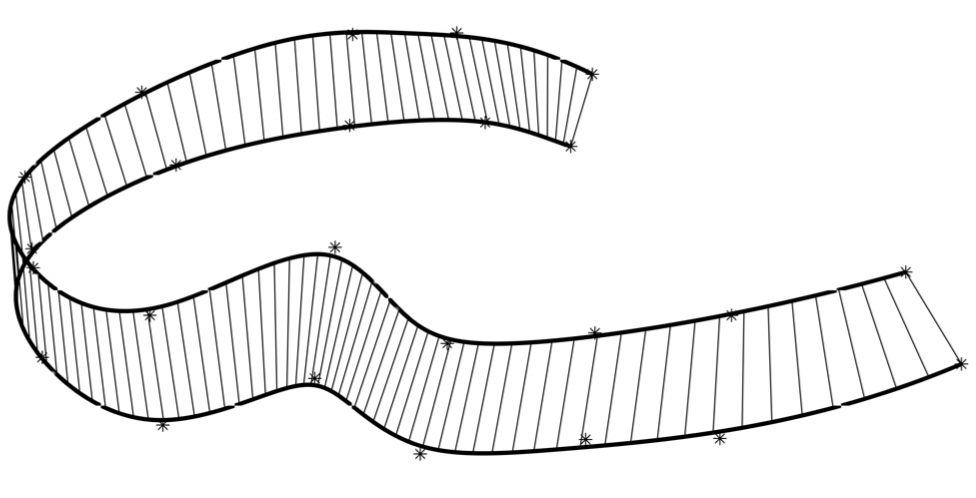}
\end{overpic}\\
\centering (e)
\end{minipage}
\hfill
\begin{minipage}[b]{0.3\linewidth}
\includegraphics[width=1\textwidth]{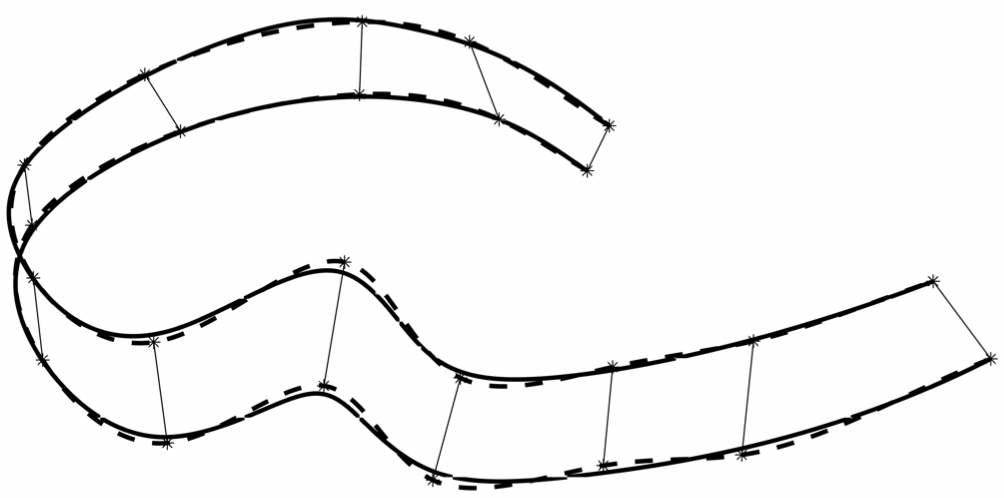}\\
\centering  (f)
\end{minipage}
\hfill
}
\centerline{
\scriptsize
\begin{minipage}[b]{0.3\linewidth}
\includegraphics[width=1\textwidth]{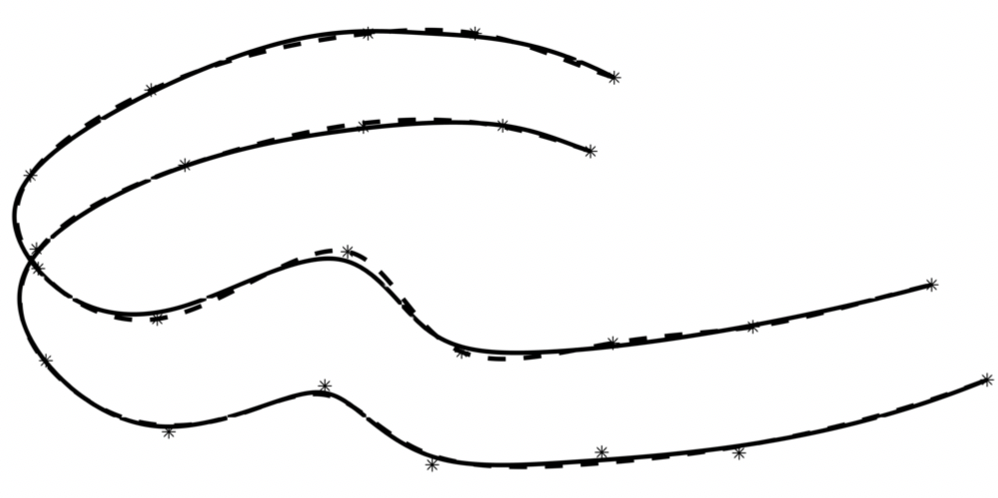}\\
\centering  (g)
\end{minipage}
}
\caption{Comparison between  the algorithms in Sec.~\ref{sec:onefixboundary} and Sec.~\ref{sec:relaxed2boundary}. $\lambda_{Energy}=0.00001, \lambda_{Width}=0.00001, \lambda_{Interior}/\lambda_{Closeness}=1$. (a) Input control rulings. (b) Initial surface $(\beta_{Max},\beta_{Ave})$=(67.52,16.81). (c) Resulting surface using the algorithm in Sec.~\ref{sec:onefixboundary} with the surface in (b) as an initialization. $(\beta_{Max},\beta_{Ave})$=(53.74,9.54), $(D_{Max},D_{Ave})$=(0.014,0.006). (d) Comparison between the initial  boundary curves in (b) (dotted curve below) and the resulting boundary curves in (c) (solid curve below). (e) Resulting surface using the algorithm in Sec.~\ref{sec:relaxed2boundary} with the surface in (b) as an initialization. $(\beta_{Max},\beta_{Ave})$=(1.28,0.28), $(D_{Max},D_{Ave})$=(0.012,0.005). (f) Comparison between initial boundary curves in (b) (dotted curve) and result boundary curves in (e) (solid curve).  (g) Comparison between resulting boundary curves in (c) (dotted curves) and (e) (solid curves).}
\label{fig:fig11}
\end{figure}

In Fig.~\ref{fig:fig11}, we  give a comparison between the performance of fixing one boundary curve (the algorithm in Sec.~\ref{sec:onefixboundary}) and relaxing both curves (the algorithm in Sec.~\ref{sec:relaxed2boundary}), with the same input control rulings, the same weight settings and termination conditions of the optimization. The results are shown in Fig.~\ref{fig:fig11}. For this experiment, adjacent lines in the input are coplanar as  shown in Fig.~\ref{fig:fig11}(a). Using the given weight setting, the algorithm fixing one boundary curve makes only a small improvement on surface developability.  The surface developability can be further improved by decreasing  $\lambda_{Interior}$,  but it will  also decrease the fitting quality. For the same input and same initial curves,  the method relaxing both boundary curves enables the resulting surface to achieve a much higher degree of developability. Moreover, the distance errors between the resulting surface and the given rulings are also smaller. 

This experiment demonstrates that  a tiny perturbation of control rulings provides a larger solution space of developable surfaces than fixing boundary curves. The optimization algorithm is thus able to find a much larger developability of the resulting surface without sacrificing the fidelity to original design intention. Since in real applications, a particular level of distance errors of  the surface is always reasonable, the proposed method of curve perturbation is a practically useful method for generating developable surfaces with a high degree of developability from control rulings.

The algorithm in Sec.~\ref{sec:onefixboundary} is also useful when fixing one boundary of the surface is a hard rule, especially when there is no strict constraint introduced by  interior rulings.  We have shown that with the first and last rulings being strict interpolation rule,  we can  still achieve a high degree of developability and the surface shape can be nicely controlled. This has been demonstrated by the results in in Fig.~\ref{fig:fig2},Fig.~\ref{fig:fig3}, Fig.~\ref{fig:fig4}, Fig.~\ref{fig:fig7}.

\subsection{Implementations}
The minimization of the nonlinear functions Eq.(\ref{eqn:F1}) and Eq.(\ref{eq:F2}) are  solved with the L-BFGS algorithm~\cite{ZHENG2012448}. The weight of each term in the objective function is determined empirically.  To unify the weight settings in the objective function, all models are scaled into a unit box. 

The developability is evaluated by the warp angle between the normal vectors at both ends of the rulings. The warp angle associated with the sampling parameter $t_i$ is defined by the angle between the vectors $N(t_i,0)$ and $N(t_i,1)$ where $N(t,s)$ is the normal vector at $S(t,s)$. Existing works have shown that the practical  requirements of ship-hull design can be met when the maximum warp angle is less than 6 degree. 

The proposed algorithms have been implemented with C++ on the Microsoft Visual Studio platform. All experiments are performed on a laptop with 2.6 GHz Intel Core i7 CPU and 16 GB 2400 MHz DDR4 memory.  The running time ranges from 16s to 60s for the experiments shown in this paper.

\section{Conclusions}

An intuitive design method is proposed to construct a quasi-developable B-spline surface through a sequence of specified rulings.  Similar to  the concept of curve design by fitting to control points, the surface is controlled by a given set of rulings.  Our algorithms have a distinct advantage over existing algorithms in that the terminal ruling lines of the resulting developable surface are specifically defined. The disadvantage of our method is that the weight settings of each term in the objective function cannot be obtained automatically. An issue worthy of further investigation is the design of the initial rulings which has an obvious impact on the degree of the developability of the resulting surface. We are also interested in the application of our methods to CNC flank  milling  since the developable surfaces can be flank-milled using conical or cylindrical cutting tools without manufacturing error.

\paragraph{Acknowledgements} This work has been supported by the National Natural Science Foundation of China (61672187) and the Key Research and Development Project of Shandong Province (2018GGX103038).


\bibliography{Developable}

\end{document}